\documentclass[manuscript]{aastex}

\newcommand{\fr}{$\nu$\ }
\newcommand{\frdot}{$\dot \nu$\ }
\newcommand{\frddot}{$\ddot \nu$\ }

\shorttitle{The Crab Pulsar Glitch of $2006$ August}
\shortauthors{Vivekanand}

\begin{document}

\title{Did the Crab Pulsar Undergo a Small Glitch in $2006$ late March/early April?}

\author{M. Vivekanand\altaffilmark{1}}

\altaffiltext{1}{No. 24, NTI Layout 1\textsuperscript{st} Stage, 3\textsuperscript{rd} Main, 
1\textsuperscript{st} Cross, Nagasettyhalli, Bangalore 560094, India.}

\email{viv.maddali@gmail.com}

\begin{abstract}
On $2006$ August $23$ the Crab Pulsar underwent a glitch, that was reported by the 
Jodrell Bank and the Xinjiang radio observatories. Neither data are available to 
the public. However, the Jodrell group publishes monthly arrival times of the Crab 
Pulsar pulse (their actual observations are done daily), using which it is shown 
that about five months earlier, the Crab Pulsar most probably underwent a small 
glitch, which has not been reported before. Neither observatory discusses the 
detailed analysis of data from $2006$ March to August; either they may not have 
detected this small glitch, or may have attributed it to timing noise in the Crab 
Pulsar. The above result is verified using X-ray data from the RXTE observatory. 
If this is indeed true, this may probably be the smallest glitch observed in the 
Crab Pulsar so far, whose implications are discussed. This work addresses the 
confusion possible between small magnitude glitches and timing noise in pulsars.
\end{abstract}


\section{Introduction} \label{sec1}

Glitches in pulsars are probably the only method of studying the internal structure of neutron
stars (\citet{Baym1969}, \citet{Ruderman1998}); see \citet{Shemar1996} and \citet{Lyne2000} 
for an observational perspective, and \citet{Haskell2015} for a theoretical discussion of 
pulsar glitches. Among pulsars that have been observed for their glitch behavior, the Crab 
Pulsar (PSR B0531+21 or J0534+2200) has been the most closely studied \citep{Lyne2015}. 
Glitches in this pulsar occur, in average, once in $1.5$ years. Its several glitches have been 
analyzed and results for them published by \citet{Lyne1988}, \citet{Lyne1993}, \citet{Lyne2000}, 
\citet{Wong2001}, \citet{Espinoza2011} and \citet{Wang2012}, the most recent study being that of
\citet{Lyne2015}. This work focuses on the glitch that occurred in the Crab Pulsar on $2006$ 
August $23$ (henceforth CPG2006); more precisely, on the five month duration before CPG2006. 
Two radio observatories observed it in sufficient detail to derive the relevant glitch parameters 
--- The Jodrell Bank Observatory (\citet{Espinoza2011}; henceforth JBO) and the Xinjiang 
Astronomical Observatory (\citet{Wang2012}; henceforth XAO). 

The critical pre-glitch reference timing model is obtained by both groups by fitting a simple 
rotation model to the timing data for a given duration before the glitch (the pre-glitch 
duration); the model typically consists of the pulsar rotation frequency \fr and its first 
two derivatives \frdot and \frddot, at the epoch of the glitch, which is MJD $\approx 53970$ 
for both observatories. The pre-glitch duration for the JBO group is about 43 days starting 
from MJD $53926$ (C. Espinoza, private communication), while that of the XAO group is $280$ 
days, from MJD $53685$ to MJD $53965$.  The timing residuals for CPG2006 relative to the 
pre-glitch reference timing model are analyzed by both groups, obtaining consistent results. 

This work shows that a different choice of the pre-glitch duration for CPG2006 
reveals what appears to be a small glitch about five months before the main glitch. 
The monthly radio timing data of the Crab Pulsar from JBO are used to derive the 
result, and X-ray data from the RXTE Observatory are used to verify it. In the last 
section it is argued that this is more likely to be a glitch than timing noise, although
the data available strictly do not allow to discriminate between the two possibilities.

\section{Observations} \label{sec2}

The Crab Pulsar was monitored daily by JBO since $1984$, mainly at $610$ MHz frequency
\citep{Lyne2015}. Occasionally it was also observed in the $1400$ to $1700$ MHz band.
The XAO has been monitoring the Crab Pulsar once a week at $1540$ MHz since $2000$ 
January \citep{Wang2012}. Both groups estimate the arrival time of the integrated pulse 
profile (IP) of the Crab Pulsar, and use this data to study the several glitches in this 
pulsar by means of standard techniques.  Neither of the above two data are available 
in the public domain. However JBO also publishes monthly arrival times of the Crab 
pulsar IP, referred to the Solar System Barycenter, and scaled to infinite frequency, in 
the so called Jodrell Bank Crab Pulsar Monthly 
Ephemeris\footnote{http://www.jb.man.ac.uk/pulsar/crab.html} (\citet{Lyne1993}; 
henceforth JBCPME). Data from this ephemeris spanning the epoch $2005$ November $15$ 
to $2007$ May $15$ (MJD $53689$ to MJD $54235$), yielding $19$ timing residuals, have 
been used for this work. In practice, one requires daily timing residuals to properly
analyze the glitches in the Crab Pulsar. However, JBCPME also contains very accurate 
\fr and \frdot at each monthly epoch, which help in converging to a sufficiently 
accurate pre-glitch reference timing model. Thus the monthly radio data used in this 
work are suitable to demonstrate the existence of the smaller glitch, which is the 
main goal of this work; derivation of statistically rigorous glitch parameters 
requires the original daily observed data. Although this work uses only the Crab 
Pulsar's timing residuals, the main result is also evident in the \fr and \frdot 
listed by JBCPME, as discussed in the last section.
\begin{figure}
\epsscale{1.0}
\plotone{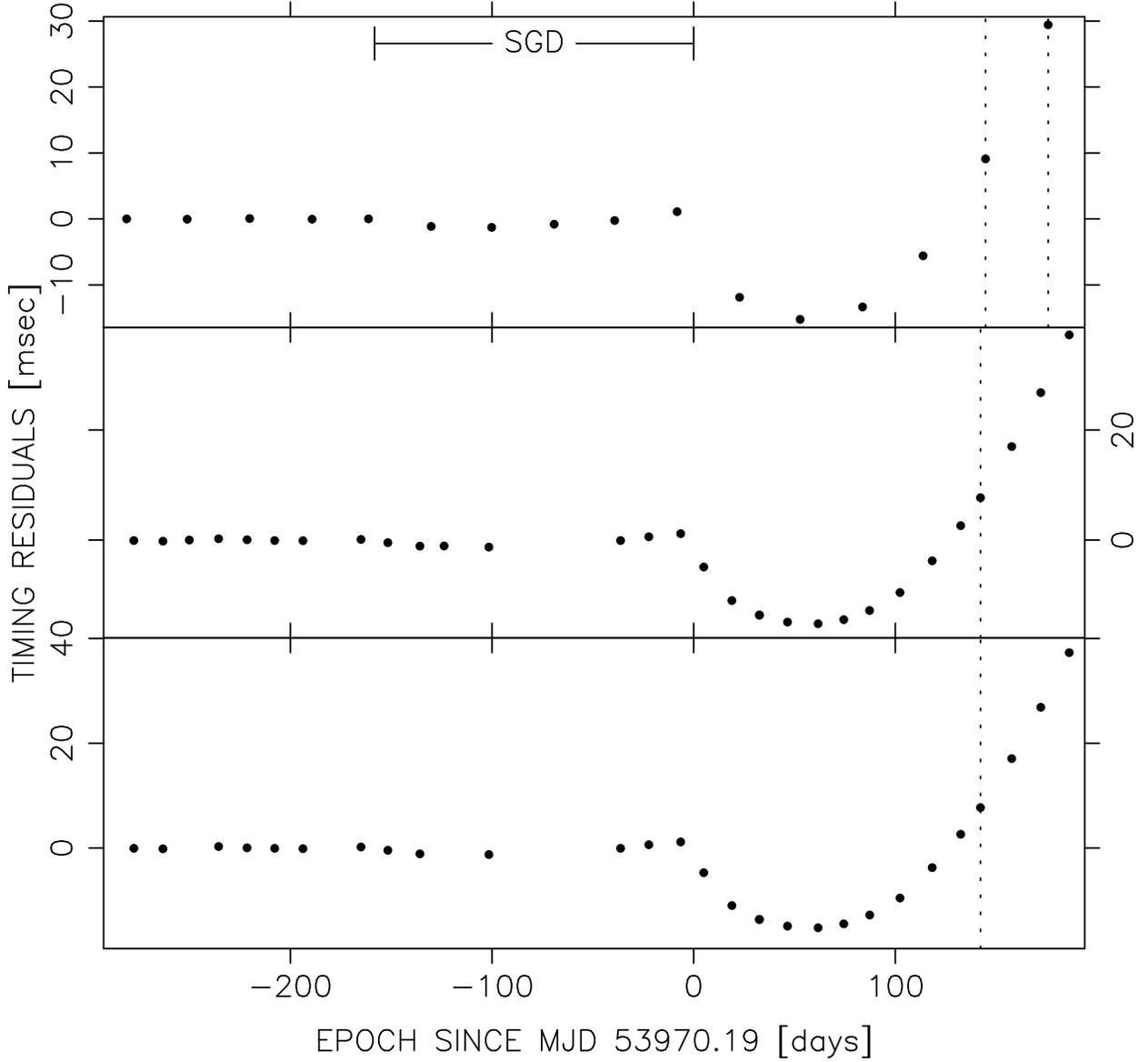}
\caption
{
Timing residuals of the Crab Pulsar relative to the pre-glitch reference timing model 
given in Table~\ref{tbl1}. Residual $0$ in each panel refers to the
mean value of the residuals belonging to the pre-glitch duration of that panel. The 
dotted vertical lines are epochs after which a phase cycle of $+1$ had to be inserted 
into the timing data (using command PHASE +1 in TEMPO$2$). Data only up to MJD 54146 
(day 175.81) has been shown to highlight the small depression extending from MJD
$\approx 53812.0$ (day $\approx -158.19$) to day MJD $53970.19$ (day $0.0$); this has 
been marked as SGD in the figure. Top panel: Radio timing residuals from JBCPME. Middle 
panel: X-ray timing residuals from the RXTE/PCA data. Bottom panel: X-ray timing 
residuals from the RXTE/HEXTE data. \label{fig1}
}
\end{figure}

The X-ray data used in this work were obtained from the Proportional Counter Array (PCA; 
\citet{Jahoda1996}) and the High Energy X-ray Timing Experiment (HEXTE; 
\citet{Rothschild1998}) aboard the RXTE observatory. The PCA consists of five 
proportional counter units (PCUs) operating in the $2$ to $60$ keV range, having a field 
of view of $1$ degree in the sky($^\circ$), and a time resolution of $1$ micro second 
($\mu$sec) (see ``The ABC of XTE'' guide on the RXTE 
website\footnote{heasarc.gsfc.nasa.gov/docs/xte/data\_analysis.html}). The HEXTE 
instrument consists of two independent clusters of detectors. Each cluster contains four 
NaI(Tl)/CsI(Na) phoswich scintillation photon counters, and has a field of view of 
$1^\circ$ in the sky. Effectively this instrument is sensitive to photons in the $15$ to 
$240$ keV range, and each photon's arrival time accuracy is $\approx 7.6$ $\mu$sec (see 
``The ABC of XTE'' guide). The first RXTE observation used in this work was obtained on 
$2005$ November $18$ (MJD $53692$), and the last on $2007$ April $23$ (MJD $54213$); 
the corresponding observation identification numbers (ObsID) are $91802$-$02$-$10$-$00$ 
and $92802$-$01$-$21$-$00$, respectively. This yielded $34$ PCA timing residuals, but 
only $32$ HEXTE residuals, since there were insufficient X-ray photons in the HEXTE 
data of ObsIDs $91802$-$02$-$12$-$00$ and $92802$-$01$-$04$-$00$. The PCA data was 
obtained in the event mode with identifier \textit{E\_$250$us\_$128$M\_$0$\_$1$s}, 
after being integrated into time intervals of $244.14$ $\mu$sec, and binned in energy 
into $128$ channels having non-uniform energy widths. In $20$ ObsIDs useful data was 
available when all $5$ PCUs were switched on; in $9$ of them when only $4$ PCUs were 
switched on; and so on. Even when only $1$ PCU was switched on (ObsIDs 
$92802$-$02$-$06$-$00$ and $92802$-$03$-$05$-$00$) the data had sufficient number of 
photons to yield a statistically significant pulse profile.

\section{Analysis of Radio Data} \label{sec3}

%
%
\begin{table}
\begin{center}
\caption{The pre-glitch reference timing model obtained using the first $5$ data in the 
top panel of Figure~\ref{fig1}, at reference epoch MJD $53750.0000002354282$ (day 
$-220.19$ in Figure~\ref{fig1}). The errors in the last digit of each number are shown 
in brackets. \label{tbl1}
}
\begin{tabular}{|c|c|}
\tableline
Parameter  & Value \\
\tableline
Epoch (MJD) & $53750.0000002354282$ \\
\tableline
$\nu$ (Hz) & $29.7749226271(1)$ \\
\tableline
$\dot \nu$ (s$^{-2}$) & $-372853.5(4) \times 10^{-15}$ \\
\tableline
$\ddot \nu$ (s$^{-3}$) & $1.2(1) \times 10^{-20}$ \\
\tableline
\end{tabular}
\end{center}
\end{table}
%
%

The top panel of Figure~\ref{fig1} shows the radio timing residuals of the Crab Pulsar, 
relative to the pre-glitch reference timing model given in Table~\ref{tbl1}, during 
the epoch under consideration. The pre-glitch reference timing model was 
obtained using the first $5$ data in the top panel of Figure~\ref{fig1}, in the range 
MJD $53673.5$ to MJD $\approx 53812.0$, translating to day $-296.69$ to $\approx 
-158.19$ in abscissa. This duration will henceforth be referred to as the pre-glitch 
duration (PGD). The TEMPO$2$ \citep{Hobbs2006} best fit parameters of this data are 
shown in Table~\ref{tbl1}.  The standard deviation of the five data, from the model 
derived by TEMPO$2$, is TRES $= 38.6$ $\mu$sec. However TEMPO$2$ estimates TRES using 
$5$ degrees of freedom, whereas the number to use is $2$,  since $3$ parameters have 
been fit for. The appropriate value of standard deviation is $\sqrt{38.6^2 * 5 / 2} = 
61$ $\mu$sec.

By including the next five timing residuals in the top panel of Figure~\ref{fig1}, 
from MJD $\approx 53812.0$ to MJD $53970.19$, or day $\approx -158.19$ to $0.0$ in 
abscissa, the TEMPO$2$ best fit parameters are $\nu = 29.7749226293(8)$ Hz, $\dot \nu 
= -372852.4(4) \times 10^{-15}$ s$^{-2}$, and $\ddot \nu = 1.14(1) \times 10^{-20}$ 
s$^{-3}$. The TRES of these ten data is $223.3$ $\mu$sec,  but the appropriate value 
of standard deviation is $\sqrt{223.3^2 * 10 / 7} = 267$ $\mu$sec. The two variances 
differ by $267^2 - 61^2 = 67568$, while the standard error on this difference is 
$\sqrt{2 \times (267^4/7 + 61^4/2)} = 38287$. The two variances differ by $67568/38287 
= 1.76$ standard errors; i.e., the latter variance is significantly larger than the 
former at the $92\%$ confidence level. Attempts made to obtain statistical fits with 
standard deviation less than $267$ $\mu$sec failed (for example, by choosing a 
different reference epoch, by altering initial values of the parameters by hand, etc). 
Therefore using radio data right up to MJD $53970.19$, to derive the pre-glitch 
reference timing model, is not justified. The reason is the small depression extending 
from MJD $\approx 53812.0$ to MJD $53970.19$ (day $\approx -158.19$ to day $0.0$) in 
the top panel of Figure~\ref{fig1}, which will henceforth be referred to as the small 
glitch duration (SGD). The main glitch that follows is clearly evident in the top panel 
of Figure~\ref{fig1}.  It starts at MJD $53970.19$ and extends up to MJD $54250.5$, or 
day $0.0$ to $280.31$ (henceforth referred to as the main glitch duration MGD), 
although data only up to MJD $54146$ (day $175.81$) has been shown in Figure~\ref{fig1} 
to highlight the depression.

\begin{figure}
\epsscale{1.0}
\plotone{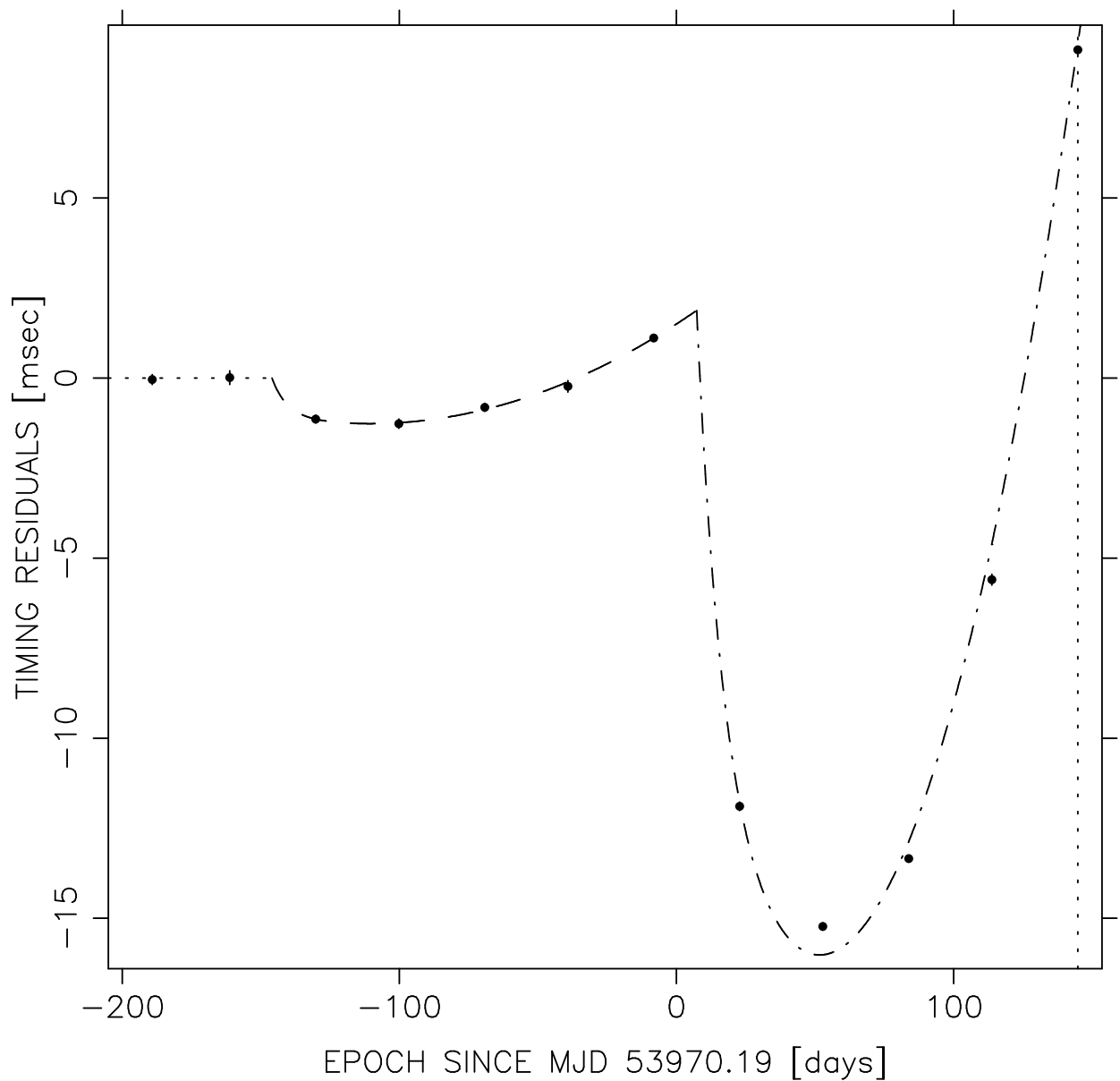}
\caption
{
A closer look at the top panel of Figure~\ref{fig1}, from MJD $53765.19$ to MJD 
$54120.19$ (days $-205$ to $+150$ in the figure). The dashed line is the function
$f(t)$, defined in Equation $1$, with the parameters given in the second row of 
Table~\ref{tbl2}. The dashed-dotted line is the function $g(t)$ with the parameters 
given in the second row of Table~\ref{tbl3}. The horizontal dotted line represents
the pre-glitch reference timing model of Table~\ref{tbl1}. \label{fig2}
}
\end{figure}

Figure~\ref{fig2} shows a closer view of the top panel of Figure~\ref{fig1}; 
earlier what looked like a depression now looks more like a glitch. Data points $3$ to 
$7$ in Figure~\ref{fig2} (five radio data points belonging to SGD) have been 
fit to the function $f(t)$ in Equation $1$, the fit taking into account their 
errors, 
\begin{eqnarray}
f(t) & = & a_1 \left (t - t_1 \right ) + \frac{b_1}{2} \left (t - t_1 \right )^2 + \nonumber \\
      &   & c_1  \left (1 - \exp \left (\frac{-(t - t_1)}{\tau_1} \right ) \right ), \nonumber \\
g(t) & = & f(t_2) + a_2 \left (t - t_2 \right ) + \frac{b_2}{2} \left (t - t_2 \right )^2 + \nonumber \\
      &   & c_2  \left (1 - \exp \left (\frac{-(t - t_2)}{\tau_2} \right ) \right ),
\end{eqnarray}
%
%
\begin{table}
\begin{center}
\caption{Results for the smaller glitch, derived from the best fit parameters,
that are obtained by fitting $f(t)$ to the data of the smaller glitch (SGD) in 
Figures~\ref{fig2} and ~\ref{fig3} (data implies timing residuals in SGD, 
relative to the model of Table~\ref{tbl1}). In all three cases, the epoch of the glitch 
$t_1$ is chosen or fixed as explained in the text. No error bars are shown for 
radio case since no degrees of freedom are left after the fit. \label{tbl2}
}
\begin{tabular}{|c|c|c|c|c|c|}
\tableline
Data  & $t_1$ (MJD) & $\Delta \nu_p$  ($10^{-6}$ Hz) & $\Delta \dot \nu_p$  ($10^{-13}$ s$^{-2}$) & $\Delta \nu_n$  ($10^{-6}$ Hz) & $\tau_d$  (days) \\
\tableline
JBCPME & $53824.2$ & $0.005$ & $-0.018$ & $0.066$ & $5.2$ \\
\tableline
PCA & $53811.9$ & $0.008 \pm 0.003$ & $-0.019 \pm 0.004$ & $0.03 \pm 0.03$ & $8.3 \pm 6.7$ \\
\tableline
HEXTE & $53811.9$ & $0.005 \pm 0.006$ & $-0.015 \pm 0.006$ & $0.03 \pm 0.04$ & $15.7 \pm 12.4$ \\
\tableline
\end{tabular}
\end{center}
\end{table}
%
%
where the parameters $t_{1,2}$ (epoch of the glitch), $a_{1,2}$ (related to the 
permanent change in rotation frequency at the epoch of the glitch $\Delta \nu_p$, 
$b_{1,2}$ (related to the 
permanent change in rotation frequency derivative $\Delta \dot \nu_p$, $c_{1,2}$ (related 
to the exponential change in rotation frequency $\Delta \nu_n$), and $\tau_{1,2}$ (decay 
time scale of the glitch) are all allowed to vary during the non-linear fit (see 
\citet{Shemar1996} and \citet{Vivekanand2015} for further details on Equation $1$). 
Several equivalent fits are obtained by choosing the initial value of $t_1$ between day 
$-160$ and day $-131$, which is the range of epoch between the last data point of PGD and 
the first data point of SGD; the corresponding $\tau_1$ obtained are $5.7$ and $3.0$ days 
respectively; larger values of $t_1$ yield smaller values of $\tau_1$, as is expected. 
An illustrative fit, obtained using the initial value of $t_1 = -146$ (midway between 
the above two numbers), is shown in the second row of Table~\ref{tbl2}. The 
corresponding $f(t)$ is plotted as the dashed line in Figure~\ref{fig2}. While the 
non-linear fit converged to the solution given with a standard deviation of $62$ 
$\mu$sec, fitting five data points to a function with five parameters leaves no 
degrees of freedom to estimate the errors on the parameters. 

\begin{figure}
\epsscale{1.0}
\plotone{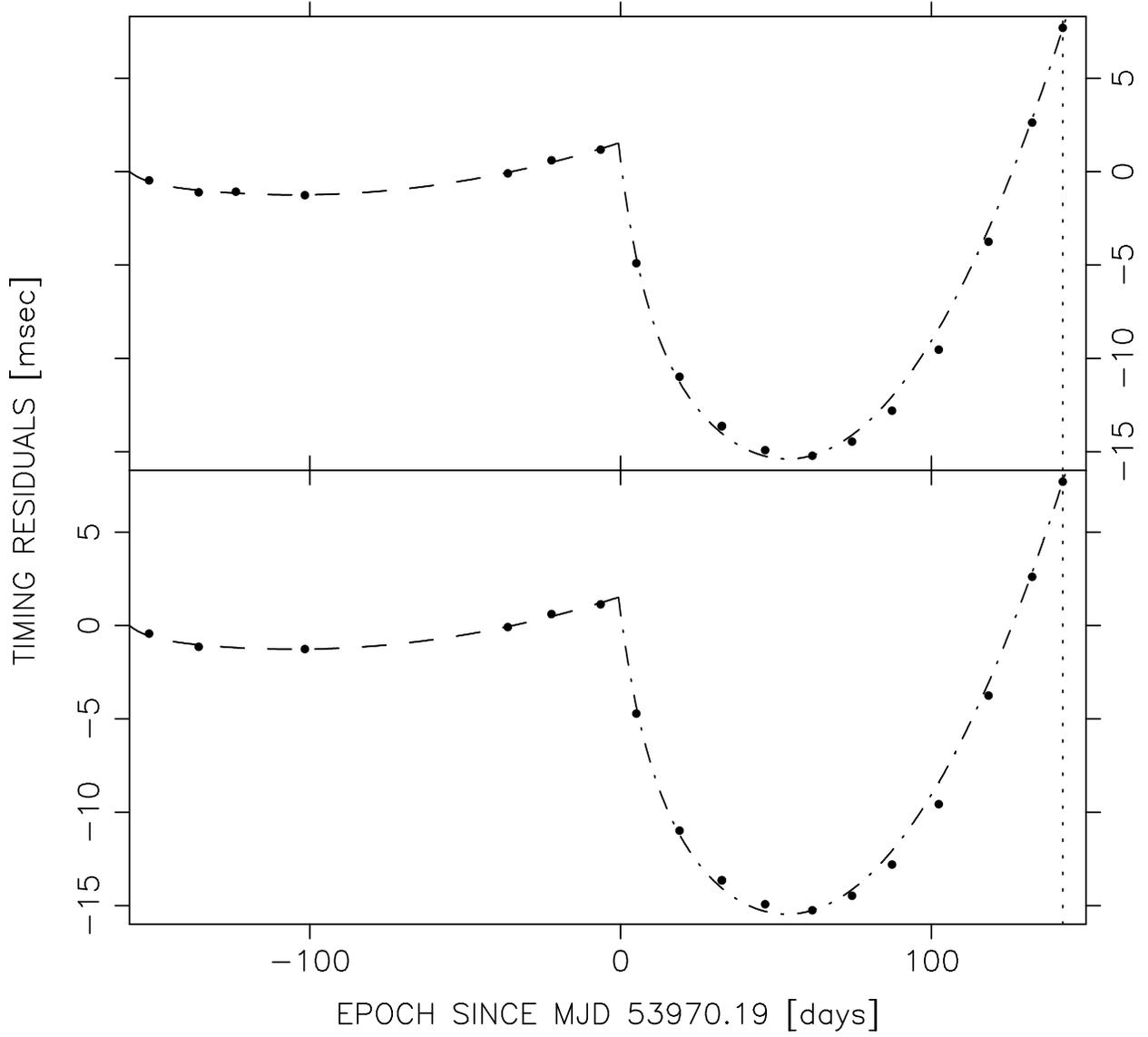}
\caption
{Top Panel: A closer look at the middle panel of Figure~\ref{fig1}, from MJD 
$53824.19$ to MJD $54120.19$ (days $-146$ to $+150$ in the figure). The dashed 
line is the function $f(t)$ with the parameters given in the third row of 
Table~\ref{tbl2}. The dashed-dotted line is the function $g(t)$ with the 
parameters given in the third row of Table~\ref{tbl3}. Bottom Panel: A closer 
look at the bottom panel of Figure~\ref{fig1}, from MJD $53824.19$ to MJD 
$54120.19$ (days $-146$ to $+150$ in the figure). The dashed line is the 
function $f(t)$ with the parameters given in the bottom row of Table~\ref{tbl2}. 
The dashed-dotted line is the function $g(t)$ with the parameters given in the 
bottom row of Table~\ref{tbl3}. \label{fig3}
}
\end{figure}

Attempts to fit the data to a modified $f(t)$, that does not contain the last ($c_1$ 
and $\tau_1$) term, converge to values of $t_1$ lower than the epoch of the last data 
point of PGD. These are unrealistic solutions, since the first residual after a glitch 
must have more negative value than that of the pre-glitch reference timing model. 
This shows that inclusion of the decay time is critical in the fit, further supporting 
the assertion that what looks like a small depression is probably a small glitch, although
timing noise can not be ruled out. However, one must keep in mind that the function $f(t)$ has five 
parameters, and only five data are used in the fit in this section. While it is true 
that a set of $5$ data points can not be fit to any arbitrary function of $5$ parameters, 
particularly if the function is a mixture of polynomials and exponential, one must be
open to the possibility that the data of SGD may also be due to timing noise.

Next, the radio data belonging to MGD are fit to the function $g(t)$ of Equation $1$. 
Ideally this should be done after subtracting $f(t)$ from the timing residuals of 
MGD. Here it is assumed that $f(t)$ is much smaller than $g(t)$ for the MGD epochs; 
so $g(t)$ is fit to the residuals without subtracting $f(t)$, since the derived
glitch parameters are not expected to be statistically rigorous anyway. It has been 
verified that results of both approaches are statistically consistent.
A fit varying all parameters converges to the unrealistic situation of $t_2$ being 
less than the epoch of the last data of SGD. By fixing $t_1$ at various epochs 
between that of last data point of SGD and that of the first data point of MGD (non 
inclusive), various solutions can be obtained in which $\tau_2$ varies consistently. 
Solutions closer to the former data point appear to have lower standard deviation. 
The solution shown in the second row of Table~\ref{tbl3} is an illustrative one, where 
$t_2$ has been fixed to a value mid way between the above two points, while the rest 
of the four parameters have been varied. The corresponding curve is the dashed-dotted 
curve in Figure~\ref{fig2}. Although the parameters in the second row of 
Table~\ref{tbl3} are consistent with the results derived by both JBO and XAO for
CPG2006, the radio data used in this work are not suitable to derive rigorous 
glitch parameters. It is 
%
%
\begin{table}
\begin{center}
\caption{Results for the larger glitch, derived from the best fit parameters, obtained by 
fitting $g(t)$ to the data of the main glitch (MGD) in Figures~\ref{fig2} and ~\ref{fig3} 
(data implies timing residuals in MGD, relative to the model of Table~\ref{tbl1}). In all 
three cases, the epoch of the glitch $t_2$ has been fixed mid way between the last data 
point of SGD and the first data point MGD.  \label{tbl3}
}
\begin{tabular}{|c|c|c|c|c|c|}
\tableline
Data  & $t_1$ (MJD) & $\Delta \nu_p$  ($10^{-6}$ Hz) & $\Delta \dot \nu_p$  ($10^{-13}$ s$^{-2}$) & $\Delta \nu_n$  ($10^{-6}$ Hz) & $\tau_d$  (days) \\
\tableline
JBCPME & $53977.5$ & $0.083 \pm 0.007$ & $-0.223 \pm 0.004$ & $0.5 \pm 0.2$ & $8.9 \pm 2.7$ \\
\tableline
PCA & $53969.5$ & $0.110 \pm 0.004$ & $-0.233 \pm 0.005$ & $0.4 \pm 0.1$ & $7.8 \pm 2.1$ \\
\tableline
HEXTE & $53969.5$ & $0.108 \pm 0.004$ & $-0.232 \pm 0.005$ & $0.3 \pm 0.1$ & $8.9 \pm 2.4$ \\
\tableline
\end{tabular}
\end{center}
\end{table}
%
%
therefore concluded that the CPG2006 event was preceded by a small glitch. However, it should be
kept in mind that this result depends critically upon the pre-glitch reference timing model derived
above, for which unfortunately only $5$ radio timing residuals were available. Therefore the 
result of this section do not rule out timing noise altogether.

\section{Analysis of X-Ray Data} \label{sec4}

The results derived using radio data are now verified using X-ray data from the RXTE 
observatory. \citet{Vivekanand2015} and \citet{Vivekanand2016} discuss in detail the 
analysis of Crab Pulsar data from the HEXTE and PCA instruments, respectively. In 
this work one has to additionally filter the PCA data for time markers, using the 
tool \textit{fselect} along with a bitfile containing the script ``Event ==  
b1xxxxxxxxxxxxxxx'' (see ``The ABC of XTE'' guide), because of the data mode of PCA.
The same must be done when using the tool \textit{seextrct} to obtain the light curve.

TEMPO$2$ fits to the $8$ PCA residuals and the $7$ HEXTE residuals in the PGD (see 
the middle and bottom panels of Figure~\ref{fig1}), resulted in pre-glitch 
reference timing models that had standard deviation of data TRES of $111$ $\mu$sec
and $114$ $\mu$sec, respectively.  After taking into account the true degrees of 
freedom, the above two standard deviations become $140$ $\mu$sec and $151$ $\mu$sec 
respectively. By including the SGD data also, the TEMPO$2$ 
fits yield pre-glitch reference timing models with corrected standard deviations of 
$231$ $\mu$sec and $268$ $\mu$sec respectively. By the argument of the previous 
section both sets of variances ($231$, $140$ and $268$, $151$) differ at the $82$\% 
confidence level. Therefore both the PCA and the HEXTE data confirm the result of 
the previous section, that one is not justified in including the SGD data to obtain 
the pre-glitch reference timing model for CPG2006.

Although the glitch behavior in the X-ray data of SGD is evident when these timing 
models are used, they are clearly worse than the pre-glitch reference timing model 
obtained from the radio data.  Therefore the latter has been used to analyze the 
X-ray data. 

The top panel of Figure~\ref{fig3} shows a closer view of the middle panel of
Figure~\ref{fig1}. Fitting the function $f(t)$ to the $7$ PCA data in the SGD 
does not converge to a solution if all five parameters are varied, for several 
initial values of $t_1$ and $\tau_1$. Therefore for illustrative purposes $t_1$ 
was fixed midway between the last PCA data point of PGD and the first PCA data 
point of SGD. The corresponding $f(t)$ is plotted as the dashed line in the top 
panel of Figure~\ref{fig3}, while the corresponding parameters, or the results 
derived from them, are listed in the third row of Table~\ref{tbl2}; the standard 
deviation of the data from this solution is $76$ $\mu$sec. Attempts to fit the 
function $g(t)$ to the $19$ PCA data of MGD by varying all five parameters either 
resulted in unrealistic solutions of $t_1$, or did not converge to a solution.
Therefore for illustrative purposes $t_1$ was fixed midway between the last PCA 
data point of SGD and the first PCA data point of MGD. The corresponding $g(t)$ 
is plotted as the dashed-dotted line in the top panel of Figure~\ref{fig3}, while 
the corresponding parameters, or the results derived from them, are listed in the 
third row of Table~\ref{tbl3}.

The bottom panel of Figure~\ref{fig3} shows a closer view of the bottom panel of
Figure~\ref{fig1}. Fitting the function $f(t)$ to the $6$ HEXTE data in the SGD 
by varying all parameters led to results similar to those in the PCA case. 
Therefore for illustrative purposes $t_1$ was fixed midway between the last HEXTE 
data point of PGD and the first HEXTE data point of SGD. The corresponding $f(t)$ 
is plotted as the dashed line in the bottom panel of Figure~\ref{fig3}, while the 
corresponding parameters, or the results derived from them, are listed in the last 
row of Table~\ref{tbl2}; the standard deviation of the data from this solution 
is $57$ $\mu$sec. Attempts to fit the function $g(t)$ to the $19$ HEXTE data 
of MGD by varying all five parameters resulted in unrealistic solutions of $t_1$. 
Therefore for illustrative purposes $t_1$ was fixed midway between the last HEXTE 
data point of SGD and the first HEXTE data point of MGD. The corresponding $g(t)$ 
is plotted as the dashed-dotted line in the bottom panel of Figure~\ref{fig3}, 
while the corresponding parameters, or the results derived from them, are listed 
in the last row of Table~\ref{tbl3}.

The behavior of the fits to X-ray data in Figure~\ref{fig3} is very similar 
to that in Figure~\ref{fig2}. It is therefore concluded that the X-ray 
data confirm the results obtained from radio data.

\section{Discussion} \label{sec5}

It is clear from the analysis of the monthly radio data (Figure~\ref{fig2}, Table~\ref{tbl2} 
and the analysis of section~\ref{sec3}) that the Crab Pulsar most probably 
suffered a small glitch before CPG2006. This is verified by the X-ray data 
(Figure~\ref{fig3}, Table~\ref{tbl2} and the analysis of section~\ref{sec4}). 

As already mentioned, the data available for this work is quite sufficient to 
demonstrate the existence of something that looks like a small glitch, but not 
to derive rigorous glitch 
parameters. So the parameters used to plot the functions $f(t)$ and $g(t)$ in 
Figures~\ref{fig2} and ~\ref{fig3} (listed in Tables~\ref{tbl2} and ~\ref{tbl3}) 
should be taken as illustrative.  Even then, the values of $\tau_2$ in 
Table~\ref{tbl3} are consistent with the values $7.3 \pm 0.3$ days derived by 
\citet{Wang2012}. The value of $\Delta \nu_p + \Delta \nu_n$ in the second row 
of Table~\ref{tbl3} is $0.58 \pm 0.20$ micro Hertz, which is consistent with the 
value of $0.41 \pm 0.09$ derived by \citet{Wang2012}. The ratio of change in 
rotation frequency $(\Delta \nu_p + \Delta \nu_n) /  \nu \times 10^9$ from the 
second row of Table~\ref{tbl3} is $19 \pm 7$, which is consistent with the value 
$21.8 \pm 0.2$ derived by \citet{Espinoza2011}. The ratio of change in rotation 
frequency derivative $(\Delta \dot \nu_p + \Delta \dot \nu_n) / \dot \nu \times 
10^3$ from the second column of Table~\ref{tbl3} is $1.8 \pm 0.7$, which is 
roughly consistent with the value $3.1 \pm 0.1$ derived by \citet{Espinoza2011}, and the 
value $1.3$ estimated from \citet{Wang2012}. Thus the illustrative radio solution 
in the second row of Table~\ref{tbl3} is consistent with earlier estimates; this
is expected since the small glitch perturbs only weakly the 
pre-glitch reference timing model for the main glitch. The 
X-ray solutions in the last two rows of Table~\ref{tbl3} are consistent with this 
radio solution. This can also be taken as indirect validation of the illustrative 
radio solution for the small glitch in Table~\ref{tbl2}, since this is used to 
derive the solution for the main glitch. The X-ray solutions in the last two rows 
of Table~\ref{tbl2} are consistent with the radio solution for the small glitch
(second row of Table~\ref{tbl2}), as expected.

The small glitch is also evident in the \fr and \frdot data of JBCPME. By fitting
a straight line to the five \frdot data of the PGD, it is seen that the next five 
\frdot (belonging to the SGD) lie systematically lower (larger in magnitude) than 
the line, by at least $-18 (\pm 7) \times 10^{-15}$ s$^{-2}$, which is the typical 
signature of a very small glitch in the Crab Pulsar. By doing the same with the \fr 
data, it is seen that \fr increases systematically in the SGD. Although one expects 
such an increase to be sudden just after a glitch, this result also indicates that 
the \fr and \frdot of the Crab Pulsar had probably behaved similar to those 
during a glitch in the boundary region between PGD and SGD.

The PGD in the radio data apparently can not be extended to lower epochs; the 
residuals for the epochs MJD $53658$ and MJD $53628$ lie systematically above
those of the five PGD residuals. This is also apparent in the \frdot data of
JBCPME, which are more positive before the PGD.

Did the glitch detector of \citet{Espinoza2014} detect this small glitch? It probably 
did, and probably classified it as one of their $381$ Glitch Candidates (GC). 
\citet{Espinoza2014} allow for the possibility that some of the GC might be real 
glitches, but believe most of them are due to timing noise. Therefore it is likely
that \citet{Espinoza2014} believe that this event is not a glitch but is on account
of timing noise.  The fact, that the $c_1$ and $\tau_1$ term in $f(t)$ (in Equation 
$1$) is required to give a sensible fit to the SGD data, gives some support for the 
belief that one is dealing with a glitch and not timing noise. On the other hand, 
\citet{Espinoza2014} may have missed this glitch, due to their 
methodology. They use $20$ times of arrival (TOA) to obtain the pre-glitch reference 
timing model, and fit a quadratic function to the next $10$ TOA. It is possible that 
these numbers are inadequate for sensing the small glitch of this work. Furthermore, 
they were looking for a sudden rise in rotation frequency $\nu$, whereas, as 
indicated by the $\nu$ data of JBCPME, this may be a slower glitch. In this context, 
this work focuses attention on the problem of distinguishing between a small glitch 
and timing noise in pulsars.

It appears unlikely that the small glitch and CPG2006 are causally connected, since they
are separated by about $\approx 150$ days. There are glitches in the Crab Pulsar separated 
by much smaller time -- the glitches of MJD $50459$ and MJD $50489$ are separated by a mere
$30$ days, although there is doubt whether the latter is a glitch at all \citep{Wong2001}.

As \citet{Espinoza2014} point out, although the exact mechanism for glitches is not
fully understood, the glitch magnitude distribution in general, and the magnitude of the smallest
glitch in particular, in the Crab Pulsar, is an important information. The magnitude of a 
glitch is related to the number of super fluid vortices that unpin during the event (see
\citet{Alpar1996} and \citet{Ruderman1998}). Clearly the smallest glitch in a pulsar will
set constraints on the minimum size of the region in the inner crust that is involved in 
the glitch process. The numbers derived in the second row of Table~\ref{tbl2} are not
reliable enough to estimate rigorously the magnitude of the small glitch. This is best done
using the daily sampled data of the JBO. The higher the dynamic range of glitch magnitudes
(the ratio of the maximum to the minimum observed glitch magnitude) in the Crab Pulsar, the 
better one can derive the distribution of glitch magnitudes; some theoretical models predict a
power law distribution \citep{Warszawski2008}.

The size of a glitch depends upon the number of unpinned vortices, the details of pinning and 
repinning of these vortices, and the location of unpining \citep{Warszawski2011}. In particular 
the size of a glitch depends upon the pinning strength, stronger pinning causing larger glitches,
although \citet{Warszawski2011} point out that this belief is  not very obvious -- contrary 
behavior is also possible. Therefore the smallest possible glitch in a pulsar may set constraints 
on the distribution of pinning strengths in the crust, at the lower end of the distribution. The
smallest possible glitch in a pulsar may also suggest that the crust is lighter rather than heavier
\citep{Warszawski2011}, which may have implications for the equation of state of a neutron star.

In summary, this work demonstrates a peculiar behavior of the timing residuals of the Crab Pulsar 
that started around late March/early April 2006, and continued up to the epoch of the main glitch.
This work has shown that this behavior is consistent with that of a small glitch. If this is true,
then this is probably the smallest glitch detected so far in the Crab Pulsar, whose implications 
have been discussed above. On the other hand this work does not rule out timing noise. Therefore,
future pulsar timing programs will need not only better sensitivity but also fast cadences in 
order to be able to study the small glitches regime. If it turns out that this is indeed timing 
noise, then there are far reaching implications for the definition of timing noise. The current 
consensus is that timing noise is supposed to be that which is left over in timing residuals 
after the effects of secular variations and glitches have been removed (\citet{Lyne1993}). Clearly 
the understanding of the term timing noise needs to be revised if it starts behaving like small glitch.

\acknowledgments

I thank the referee Crist\'obal Espinoza for detailed and helpful comments.
This research made use of data obtained from the High Energy Astrophysics Science 
Archive Research Center Online Service, provided by the NASA-Goddard Space Flight 
Center.

\end{document}